# The Description and Scaling Behavior for the Inner Region of the Boundary Layer for 2-D Wall-bounded Flows

By David W. Weyburne[1]
*Air Force Research Laboratory, 2241 Avionics Circle, Wright-Patterson AFB, OH 45433, USA*

## Abstract

A second derivative-based moment method is proposed for describing the thickness and shape of the region where viscous forces are dominant in turbulent boundary layer flows. Rather than the fixed location sublayer model presently employed, the new method defines thickness and shape parameters that are experimentally accessible without differentiation. It is shown theoretically that one of the new length parameters used as a scaling parameter is also a similarity parameter for the velocity profile. In fact, we show that this new length scale parameter removes one of the theoretical inconsistencies present in the traditional Prandtl Plus scaling's. Furthermore, the new length parameter and the Prandtl Plus scaling parameters perform identically when operating on experimental datasets if the Rotta similarity constraint ($u_\tau/u_e$ = constant) holds. This means that many of the past successes ascribed to the Prandtl Plus scaling also apply to the new parameter set but without one of the theoretical inconsistencies. Examples are offered to show how the new description method is useful in exploring the actual physics of the boundary layer.

## 1. INTRODUCTION

The turbulent boundary layer formed by fluid flow along a wall has been an active area of research since the concept of boundary layers was introduced by Prandtl more than one hundred years ago. Early on it was realized that the turbulent boundary layer consisted of two distinct regions. A region close to the wall where viscous effects are important and an outer region where viscous effects are mostly absent. More recent research divides the turbulent boundary layer even further with the addition of other regions (see, for example, George[1] and Klewicki, *et al.*[2]). One of these added regions includes the so-called Logarithmic Law region located in between the inner and outer regions. In spite of much research in this area, the exact locations and extent of these regions have never been firmly established. What has been done in the past is to examine some high-quality experimental datasets and visually assign approximate start and stop locations to the different regions. These assigned boundary locations are then assumed to apply to all wall-bounded turbulent boundary layers. The underlying assumption is that the velocity profiles of the inner viscous region and the Logarithmic Law region all behave similarly with scaled with the Prandtl Plus scaling parameters. That is, if one plots all of the scaled velocity profiles for all wall-bounded turbulent boundary layers on one graph, the profiles will all plot on top of one another from the wall to the outer extent of the Logarithmic Law region. By assigning the region boundaries in terms of the Prandtl Plus length scale, then one supposes that these assignments are universal for all wall-bounded turbulent boundary layers.

[1]David.Weyburne@gmail.com

However, this universal picture of the wall-bounded turbulent boundary layers has come under increased scrutiny. For example, it appears that the low Reynolds number turbulent flows do not even have a Logarithmic Law region. Then there is the on-going debate as to what is the proper start and end extent of Logarithmic Law region (see Marusic, *et al*.[3]). The large scatter in the assignment of the start and stop locations from the literature brings into question the universal picture. Even more damaging is the recent revelations by Weyburne[4,5] that cast serious doubt on the universality of similarity of the viscous region using Prandtl Plus scaling's. In one paper, Weyburne[4] used the flow governing equation approach to similarity and found that the ONLY situation in which the Prandtl Plus parameters are possible similarity scaling parameters is for the laminar and turbulent sink flow cases. The Prandtl Plus scaling's failed for the more general Falkner-Skan flows. In a second paper, Weyburne[5] proved that the Prandtl Plus scaling's are ONLY strictly valid for velocity profile data sets displaying whole profile similarity. The problem, of course, is that whole profile similarity, with the exception of sink flow, of the wall-bounded turbulent boundary layer has NEVER been seen using any set of scaling parameters.[6,7] The bottom line is that this picture of a boundary layer where the different regions are located at fixed locations is probably not viable and in fact is obstructing the discovery of the true physics of the boundary layer.

If we are to move away from a universal fixed boundary picture of the turbulent boundary layer then what is needed is a way to actually physically describe and measure these regions based on experimentally accessible information. What we propose herein is to use descriptive parameters based on information extracted from the second derivative of the velocity profile. It is important to point out that these parameters can be extracted from experimental profiles without having to actually differentiate the velocity profile. The second derivative is chosen as a measure of the extent of the viscous forces in the boundary layer. The Navier-Stokes equation that describes the momentum transport in the boundary layer tells us that the viscosity influences are important where the second derivative of the velocity (second derivative with respect to the wall-normal coordinate) is significant. In order to define the thickness and shape of the viscous boundary layer region, we propose that the integral moment method developed by Weyburne[8,9] can be used. This method is based on the same integral moment type description used to characterize probability distribution functions and was developed from the realization that the second derivative of the Blasius velocity profile solution looks very much like a Gaussian probability distribution. Using this second derivative-based moment method, we believe that it can serve as the basis for describing the inner region of the turbulent boundary layer, including the Logarithmic Law region.

An advantage of this new integral based inner region description is that it can be proven (Section 3 below) that the new length scale is also a similarity length scale. That is, if similarity is present in a set of velocity profiles then the new inner region length scale must be a similarity length scale. One might suppose that this is an unlikely event given that whole profile similarity of the wall-bounded turbulent boundary layer has never been seen using any set of scaling parameters.[6,7] However, while strict whole profile similarity is unlikely, it appears that almost similar-like conditions prevail in many inner regions of turbulent boundary layer flows. This seems to be the case for many datasets using the Prandtl Plus scaling's. In what follows, we point out that plots of experimental datasets using the new parameters as scaling parameters will result in the same exact similar-like behavior as the Prandtl Plus scaling's. The advantage of the new scaling's, at least in this regard, is that they do not suffer from the Falkner-Skan failure[4] that is present in the Prandtl Plus scaling's.

In what follows, we start by reviewing the new inner region scaling parameters. We then go on to show how this new scaling is also a similarity length scale.

## 2. The Inner Region Definition

The new boundary layer thickness description[8-9] employs an integral moment-based method of the velocity along the wall and its first two derivatives in the direction normal to the plate. It is our contention that the inner region of the turbulent boundary layer can be described in terms of the second derivative-based formulation. The reason is simple; the viscous term of the Navier-Stokes momentum equation that dictates flow behavior along the wall is directly proportional to the second derivative of the velocity. Hence, by using the second derivative profiles we are defining the region where viscous forces are important. To put the new formulation into proper perspective, we first briefly review the relevant equations for this inner region description.

The mathematical description of the inner layer region of the turbulent boundary layer borrows from probability density function methodology and is based on central moments of the second derivative-based kernel. For wall-bounded 2-D flow along a flat wall, let $y$ be the normal direction to the wall, $x$ is the flow direction, and $u(x,y)$ is the velocity parallel to the wall in the flow direction. We define the viscous velocity boundary layer $n$th central moment, $\lambda_n$ as

$$\lambda_n(x) \;=\; \int_0^h dy\,\left(y-\mu_1(x)\right)^n \frac{d^2\left\{-\mu_1(x)\,u(x,y)/u_e(x)\right\}}{dy^2} \quad , \tag{1}$$

where $y=h$ is deep into the free stream, $u_e(x)$ is the velocity at the boundary layer edge, and where the first moment about zero, $\mu_1(x)$, is also the normalizing parameter. The derivative in Eq. 1 is written in this way to emphasize the probability-density-function-like behavior. It is straightforward to show that

$$\mu_1(x) \;=\; \frac{u_e(x)}{\left.\dfrac{du(x,y)}{dy}\right|_{y=0}} \;=\; \frac{\nu u_e(x)}{u_\tau^2(x)} \quad , \tag{2}$$

where $u_\tau(x)$ is the friction velocity, and $\nu$ is the kinematic viscosity. The first moment about the origin, $\mu_1(x)$, will be referred to as the viscous mean location. This is a characteristic parameter of the any boundary layer region where viscous forces are present. An important attribute of this parameter is that it is experimentally accessible without differentiation as long as $u_e(x)$ and $u_\tau(x)$ are known.

The second central moment is related to a parameter we call the viscous boundary layer width given by $\sigma_v=\sqrt{\lambda_2}$. Using integration by parts, this parameter can be shown to reduce to $\sigma_v=\sqrt{2\mu_1\delta_1-\mu_1^2}$, where $\delta_1$ is the displacement thickness. This makes this parameter experimentally accessible without differentiation as long $\delta_1(x)$, $u_e(x)$ and $u_\tau(x)$ are available. The physical description of the shape of the viscous boundary layer is extended by using the third and fourth moments to define the viscous boundary layer skewness $\gamma_{1v} = \lambda_3 / \sigma_v^3$ and the viscous boundary layer excess $\gamma_{2v} = \lambda_4 / \sigma_v^4 - 3$. These parameters can also be calculated without differentiation using integration by parts. The resulting expressions[9] are $\lambda_3 = 2\mu_1^3 - 6\mu_1^2\alpha_0 + 6\mu_1\alpha_1$ and $\lambda_4 = -3\mu_1^4 + 12\mu_1^3\alpha_0 - 24\mu_1^2\alpha_1 + 12\mu_1\alpha_2$ where $\alpha_n$ are displacement like integrals given by

$$\alpha_n = \int_0^h dy\, y^n \left(1 - u(x,y)/u_e(x)\right) \quad . \tag{3}$$

The definition of the viscous boundary layer thickness $\delta_v$, defined as the point at which the viscous contributions to the stream-wise velocity component becomes negligible, is taken as $\delta_v = \mu_1 + 2\sigma_v$. It should be noted that the probability community sometimes uses the mean plus three or four times $\sigma_v$ instead of two-sigma as used herein. Any of these definitions can be used as long as it is clear which of the definitions is being used. The two-sigma value was chosen so that $\delta_v$ approximately tracks the 99% thickness $\delta_{99}$ for laminar flow. This parameter is experimentally accessible as long $\delta_1(x)$, $u_e(x)$ and $u_\tau(x)$ are known.

## 2.1 Experimental Application

To demonstrate the new descriptive parameters, consider six second derivative profile curves from Sillero, Jimenez, and Moser[10] plotted in Fig. 1. The curves are generated by 2-pnt numerical differentiation of the velocity profile data. The two parameters $\mu_1$ and $\sigma_v$ were calculated in two ways. In the first case, we use the actual numerical derivative data and then numerically integrate the first moment about zero $\mu_1$ and the second central moment $\lambda_2 = \sigma_v^2$. In the second case, for $\mu_1$, we use Eq. 2 and author supplied $\nu$, $u_\tau$, and $u_e$ data. For the second case for $\sigma_v$, we calculated its value using $\sigma_v = \sqrt{2\mu_1\delta_1 - \mu_1^2}$ where the displacement thickness $\delta_1$ is calculated from the velocity profile data. The calculated values for $\mu_1$ and $\sigma_v$ for the two cases are found to be numerically very close (within ~0.1%). This result is important in that it verifies that these parameters can be obtained from readily available experimental measurements without having to differentiate the velocity profile.

There are two significant takeaways from the Fig. 1 graph. First, notice the location of $\mu_1^+$ in Fig. 1. The mean location is not near the peak value as one might expect from a Gaussian curve but is very far off to the side of the peak. The reason for this is the very slow decay of the peaks tail region. The value for $\sigma_v^+$ is very large (see Fig. 1 inset), again attributable to the slow peak decay. Calculating the second derivatives thickness value as $\delta_v^+ = \mu_1^+ + 2\sigma_v^+$ gives a value for all six profiles over 220 in Plus units. Recall that in the past the viscous effect on the turbulent boundary layer is considered to only extend to $y^+ \cong 30$.[1] The actual thickness $\delta_v^+$ of the viscous region is more than eight times this value (see Fig. 1 inset)! The point to be made is that the viscous region extends much farther into the fluid than previously believed due to the slow peak decay.

A second important takeaway from Fig. 1 is the behavior of the $\mu_1$ and $\sigma_v$ parameters as a function of Reynolds number. It is apparent that $\mu_1$ and $\sigma_v$ calculated in Prandtl Plus units are increasing with Reynolds number (see Fig. 1 inset). One may be tempted to claim that the inner region shows velocity profile similarity by looking at the graph. All of the profiles basically plot on top of one another. However, looking at the $\mu_1^+$ and $\sigma_v^+$ values tells a different story. Although the $\mu_1^+$ values are close in value, they are increasing with Reynolds number as are $\sigma_v^+$ and $\delta_v^+$. This means the viscous region size is changing with Reynolds number when plotted in Prandtl Plus scaling. If similarity was indeed present, all of the $\mu_1^+$ values and all of the $\sigma_v^+$ values would be equal. So, although of the graph seems to show similar behavior, the six profiles inner regions are NOT similar using the Prandtl Plus scaling's.

### 2.2 Viscous Region Extent

The next question to consider is the extent of this new viscous region definition. The parameter $\delta_v$ is a measure of the outer boundary of the viscous region. What I not clear is how this new description compares to the sublayer model currently in use. This is most easily seen by examining some second derivative profiles generated from velocity profile data from the literature. In Figs. 2a-4a we show some examples from the literature using additional datasets from Österlund[11] and DeGraaff and Eaton.[12] What is easily noticed is that the tail region of the second derivative profiles is falling off roughly as $1/(y^+)^2$ (second derivative of the Logarithmic Law of the Wall). Hence, the new viscous region model starts at the wall and extends to the outer extent of Logarithmic Law region. According to classical theory, this outer bound is a fixed fraction of the boundary layer thickness, $\delta$, with typical estimates ranging from $0.1\delta$ to $0.2\delta$.[3] In the new definition, it is simply given by the $\delta_v$ value measured for each dataset.

## 3. Similarity of the 2-D Wall-bounded Flow

In the last section, a new way to describe the inner region for the wall-bounded turbulent boundary layer was outlined. In this Section we prove that one of the new thickness parameters must also be similarity length scale. This is done by starting with the definition of similarity. For 2-D wall-bounded flows, velocity profile similarity is defined as the case where two velocity profiles taken at different stations along the flow differ only by simple scaling parameters in $y$, the normal direction to the wall, and $u(x,y)$, the velocity parallel to the wall in the flow direction. According to Schlichting,[13] a velocity profile at position $x_i$ is said to be similar to the velocity profile at $x_j$ if

$$\frac{u\left(x_i, y/\delta_s(x_i)\right)}{u_s(x_i)} = \frac{u\left(x_j, y/\delta_s(x_j)\right)}{u_s(x_j)} \quad \text{for all y,} \tag{4}$$

where the length scaling parameter is $\delta_s(x)$ and the velocity scaling parameter is $u_s(x)$. Although the intent of Schlichting's Eq. 4 expression is clear, it is based on redefining the velocity function $u(x,y)$. The mathematically correct definition is to express velocity profile similarity as

$$\frac{\overline{u}(x_i,y_i)}{u_s(x_i)} = \frac{\overline{u}(x_j,y_j)}{u_s(x_j)} \quad \text{where} \quad \overline{u}(x_i,y_i)=u(x_i,y), \quad y_i = y_j, \quad \text{and} \quad y_i=\frac{y}{\delta_s(x_i)} \qquad \text{for all } y. \tag{5}$$

For the work herein, we will use Eq. 5 as the operational definition for velocity profile similarity.

If one plots the scaled velocity $\overline{u}/u_s$ versus the scaled height $y/\delta_s$ for a set of profiles displaying similarity, then the profiles will plot on top of one another. The new similarity approach is based on a simple concept starting with this idea and that is that the area under this set of scaled velocity profile curves that show similar behavior must be equal. This can be taken a step further; if we multiply both sides of Eq. 5 by $y_i^n$, then both sides must still be equivalent. If we then integrate $y_i^n$ times both sides with respect to $y_i$ then they are still equivalent and we end up with moment-like integrals. As we will demonstrate below, the moment-like integrals provide a way of discovering different properties inherent to velocity profile similarity. In fact, this method has been used to show that if similarity is present in a set of velocity profiles, then the displacement thickness, $\delta_1$, must be a similarity length scale.[14] This method applies to any 2-D wall bounded set of profiles displaying similarity.

### 3.1 Second Derivative Profile Similarity

If similarity is present in a set of velocity profiles, then it is self-evident that the properly scaled second derivative profile curves (second derivative of Eq. 5 with respect to $y_i$) must also be similar. It is also self-evident that the area under the plotted scaled second derivative profile curves must be equal for similarity. In mathematical terms, the area under the scaled second derivative profile curve, $a(x_i)$, is expressed by

$$\begin{aligned}
a(x_i) &= \int_0^{\overline{h}} dy_i \; \frac{d^2\left\{\overline{u}(x_i,y_i)/u_s(x_i)\right\}}{dy_i^2} \\
a(x_i) &= \frac{1}{u_s(x_i)}\int_0^{\overline{h}} d\left\{\frac{y}{\delta_s(x_i)}\right\} \frac{d^2\overline{u}(x_i,y_i)}{d\left\{\dfrac{y}{\delta_s(x_i)}\right\}^2} = \frac{\delta_s(x_i)}{u_s(x_i)} \int_0^{h_i} dy \; \frac{d^2u(x_i,y)}{dy^2} \\
a(x_i) &= \frac{\delta_s(x_i)}{u_s(x_i)}\left[\frac{du(x_i,y)}{dy}\right]_{y=h_i,0} \\
a(x_i) &= -\frac{\delta_s(x_i)}{u_s(x_i)}\left[\frac{du(x_i,y)}{dy}\right]_{y=0} ,
\end{aligned} \tag{6}$$

where $\overline{h}=h_i/\delta_s(x_i)$ is located deep into the free stream. The velocity derivative at the wall is defined in terms of the friction velocity as

$$\left.\frac{du(x,y)}{dy}\right|_{y=0} = \frac{u_\tau^2(x)}{\nu} . \tag{7}$$

As long as $h_i / \delta_s(x_i) = h_j / \delta_s(x_j)$, then a necessary but not sufficient condition for similarity is that $a(x_i) = a(x_j)$. Combining Eqs. 2, 6, and 7, it is apparent that similarity requires

$$\frac{\delta_s(x_i)}{u_s(x_i)} \frac{u_e(x_i)}{\mu_1(x_i)} \quad = \quad \frac{\delta_s(x_j)}{u_s(x_j)} \frac{u_e(x_j)}{\mu_1(x_j)} \quad = \quad \text{constant} \quad . \tag{8}$$

The importance of this equation is that if similarity is present in a set of velocity profiles for any 2-D wall bounded flow, then the ratio of the scaling parameters must be proportional to the ratio of the velocity at the boundary layer edge to the mean location.

Next, consider what happens when we take the second derivative of both sides of Eq. 5 and then multiply both sides by $y_i$. The equivalence condition still holds. If we integrate the result, then the area under $y_i$ times the scaled second derivative profile curve, $b(x_i)$, is expressed by

$$\begin{aligned}
b(x_i) \quad &= \quad \int_0^{\bar{h}} dy_i \;\; y_i \frac{d^2 \left\{ \bar{u}(x_i, y_i) / u_s(x_i) \right\}}{dy_i^2} \\
b(x_i) \quad &= \quad \frac{1}{u_s(x_i)} \int_0^{\bar{h}} d\left\{ \frac{y}{\delta_s(x_i)} \right\} \; \frac{y}{\delta_s(x_i)} \frac{d^2 \bar{u}(x_i, y_i)}{d\left\{ \dfrac{y}{\delta_s(x_i)} \right\}^2} \quad = \quad \frac{1}{u_s(x_i)} \int_0^{h_i} dy \;\; y \frac{d^2 u(x_i, y)}{dy^2} \\
b(x_i) \quad &= \quad \frac{1}{u_s(x_i)} \left[ y \frac{du(x_i, y)}{dy} \right]_{y = h_i, 0} - \frac{1}{u_s(x_i)} \int_0^{h_i} dy \;\; \frac{du(x_i, y)}{dy} \\
b(x_i) \quad &= \quad - \frac{1}{u_s(x_i)} \left[ u(x_i, y) \right]_{y = h_i, 0} \quad = \quad - \frac{u_e(x_i)}{u_s(x_i)} \quad .
\end{aligned} \tag{9}$$

If the two profiles are similar at $x_i$ and $x_j$, then we must have $b(x_i) = b(x_j)$. Hence, if similarity is present in a set of velocity profiles, then the velocity scaling parameter $u_s(x)$ must be proportional to the velocity at the boundary layer edge, $u_e(x)$. This result was obtained earlier by Weyburne[14] using the equal area similarity method and by Rotta[15] and Townsend[16] using the flow governing equation approach for turbulent profile similarity.

If we combine Eqs. 8 and 9, then it is evident that

$$\frac{\delta_s(x_i)}{\mu_1(x_i)} \quad = \quad \frac{\delta_s(x_j)}{\mu_1(x_j)} \quad = \quad \text{constant} \quad . \tag{10}$$

This means that a necessary but not sufficient condition for similarity is that the mean location, $\mu_1(x)$, must be a similarity length scale.

## 4. Discussion

The traditional description of the inner region of the turbulent boundary layer starts out with a linear region near the wall, then a buffer region between the linear and the logarithmic regions, and finally the logarithmic region.[1] Notice this description is actually describing the behavior of the velocity profile. What is lost in this description is what is happening with the physics of the flow in these sub-regions. We know the viscous forces are important in the inner region but it is not obvious what role they play in the sub-regions of the traditional model. The traditional description is, in fact, obscuring the role instead. Consider the so-called linear region, the region where the profile behaves linearly. If the profiles really behaved linearly, the viscous forces would be zero in this sub-region (the second derivative of the velocity would be zero). In fact, the opposite is true, the viscous forces actually peak in this sub-region. If one combines this murky description with the problems already discussed above concerning the fixed boundary extents using Prandtl Plus scaling, then we have a traditional description that is less than optimal. On the other hand, the results above make it clear that the integral moment method for describing the thickness and shape of the inner region of the wall-bounded turbulent boundary layer has a number of advantages compared to the fixed boundary method presently employed. First, there is the fact that the second derivative moments are part of a larger moment method for describing the whole boundary layer region.[9] Next, consider the theoretical advantages. The new viscous mean location parameter must be a similarity scaling parameter and it does not suffer the Prandtl Plus scaling failure[4] (see Appendix). Perhaps most importantly, the new descriptive parameters are experimentally accessible. This means we can explore what is happening with the viscous forces in the boundary layer for each dataset using readily available measured parameters.

The importance of being able to directly explore what is going on in the viscous region is demonstrated in Figs. 1 and 2. In those figures, we show that the appearance of similarity in Sillero, Jimenez, and Moser's[10] dataset, when plotted with the Prandtl Plus scaling (Fig. 1), is just that, similarity in appearance only. When the characterizing parameters $\mu_1^{\ +}$ and $\sigma_v^{\ +}$ are examined, they show a steady increase with Reynolds number indicating that the dataset is not showing true similarity of the inner region using the Prandtl Plus scaling's. Another example of the importance of being able to measure and characterize the viscous region directly was revealed by Weyburne[9] in which he demonstrated how the $\mu_1$ parameter tracks the laminar to turbulent boundary layer transition and identifies the exact point where transition to fully turbulent flow occurs. A third example concerns the datasets from Österlund[11] and DeGraaff and Eaton[12] depicted in Figs. 3a and 4a. In the figure insets, we show the calculated values for $\mu_1$, $\sigma_v$, and $\delta_v$ in millimeters. What is interesting about this data is that they both show that the viscous region is actually shrinking with increasing Reynolds number. What distinguishes these two zero pressure gradient (ZPG) datasets is that the velocity profile measurements were not done by taking profiles at different points along the wall but were taken by sitting at one location and changing either the inlet velocity or the kinematic viscosity (or both). Interestingly, when we compare these fixed location datasets to conventional datasets that were generated by moving along the plate, we find a marked difference. For conventional ZPG cases from Wieghardt and Tillmann[17] and Smith,[18] what we observe is that $\mu_1$, $\sigma_v$, and $\delta_v$ values (see Table 1 and 2) all increase with Reynolds number indicating the viscous region is expanding with Reynolds number. This means that the velocity profile datasets taken as a function of Reynolds number for the two experimental methods are NOT equivalent. This reinforces the growing literature (see Marusic, *et al*.[21] and references therein) that all ZPG datasets are not equivalent. These examples make it clear that if this new description method was widely applied, then it could be a powerful step forward in helping to understand the physics of the wall-bounded turbulent boundary layer.

The similarity results in Section 3 indicate that if similarity is present in a set of velocity profiles, then $\mu_1(x)$ and $u_e(x)$ must be similarity scaling parameters. One of the advantages of the new scaling's is that they do not suffer from the Falkner-Skan failure[4] that is present in the Prandtl Plus scaling's (see Appendix). However, based on the wall shear stress argument,[5] both parameter sets are only strictly valid if whole profile similarity is present. Earlier we noted that strict whole profile similarity has not been observed for turbulent boundary layers.[6,7] Even though strict whole profile similarity does not seem to be present, there are many datasets in the literature that show almost similar behavior using the Prandtl Plus scaling's. This is apparent from our discussion of Fig. 1 above. The six profiles show similar-like behavior but the measured thicknesses indicate that they are not truly similar. Other examples are shown in Figs. 3b and 4b. The velocity profile datasets all look similar-like in the inner region. The bottom line is that apparently, similar-like behavior is possible so that it may be that this same type of behavior for experimental datasets can also happen for the $\mu_1(x)$ and $u_e(x)$ scaling parameters.

To verify this almost similar behavior using the new parameter set we could replot the datasets using the new parameter set to see whether they work as well as the Prandtl Plus set. However, it turns out that it is not necessary to replot the experimental profiles to do this. If one looks at how turbulent boundary layer velocity profile similarity is normally checked, then the new scaling parameter set will actually perform identically to the Prandtl Plus scaling's for experimental data sets if the Rotta[15] similarity constraint ($u_\tau/u_e$ = constant) holds. This is because what is traditionally done to discover similar behavior is to plot all of the scaled velocity profiles on one graph and use "chi-by-eye" to decide whether similarity is present. If we use Eq. 2 for $\mu_1(x)$, then examination of the scaled *x* and scaled *y* variables for the two cases indicates that the scaled profile plots will look identical. The key to this is realizing that multiplying/dividing both the scaled *x* and scaled *y* variables by common factors, in this case $u_\tau / u_e$, keeps the relative relationships between the curves the same. The actual *x* and *y* scales will be different but the plotted curves relative to each other will be identical. Hence, for experimental data sets, the two plotted parameter sets will look and behave identically. This means that all of the experimental results obtained for the Prandtl Plus scaling's that show good curve overlap will also apply to the new scaling's. Therefore, the new parameter scaling set $\mu_1(x)$ and $u_e(x)$ will have the same experimental results as the Prandtl Plus scaling but without the theoretical problems.

This experimental observation may explain the apparent success of the Prandtl Plus scaling for the last one hundred some years. What we would put forth is that it is not the Prandtl Plus scaling that has succeeded in the past but the $\mu_1(x)$ and $u_e(x)$ set instead. According to the standard method of plotting all of the scaled profiles on one graph, the two sets are the same from an experimental plot standpoint. However, theory tells us that the Prandtl Plus scaling fails[4] for the general Falkner-Skan type power law boundary layer flow but the $\mu_1(x)$ and $u_e(x)$ set works just fine (see Appendix). Theory tells us that if similarity is present in a set of profiles then the $\mu_1(x)$ and $u_e(x)$ parameter set must be similarity scaling parameters. The Prandtl Plus scaling parameters only work as similarity scaling parameters for sink flows.[4] As a length scale, $\mu_1(x)$ has a solid theoretical foundation as the first moment about zero and is part of a robust system[9] of defining the length and shape of the whole boundary layer. As a length scale, Prandtl Plus length scale has no similar theoretical foundation. The velocity scaling parameter $u_e(x)$ has been proven to be a similarity scaling parameter using both the equal area approach (Eq. 8 and ref. 14) and the flow governing equation approaches.[1,15,16] This is not true for the Prandtl Plus scaling velocity parameter. Therefore, it may be that all of the past success of the Prandtl Plus scaling is actually due to the success of the $\mu_1(x)$ and $u_e(x)$ parameter set.

Recall that in a previous paper Weyburne[4] introduced an alternative inner region scaling parameter set based on a Falkner-Skan similarity approach. It is worth discussing how the new alternative parameter set combination $\mu_1$ and $u_e$ differs from the Falkner-Skan alternative set. In the earlier work, two new scaling parameters $\delta_0$ and $u_0$ were introduced such that their ratio was directly proportional to the friction velocity squared. This by itself is not enough to fully define the parameters. To complete the definition, Weyburne[4] then assumed that the resulting Falkner-Skan $\alpha$ and $\beta$ expressions must be constants. By assuming power law expressions for $u_e$ and $u_\tau$ in terms of the distance along the wall in the flow direction, one can then find power-law expressions for $\delta_0$ and $u_0$. These expressions turn out to be identical to Falkner-Skan[20] original parameter scaling expressions. The power law expressions for $\delta_0$ and $u_0$ were shown to produce similarity-like behavior in the inner region for a number of experimental turbulent boundary layer datasets.[4] This experimental success would also apply to the new $\mu_1(x)$ and $u_e(x)$ parameter set given by Eq. A12 (Appendix). What is important to consider here is that the Falkner-Skan approach does not identify the actual scaling parameters, but rather the functional behavior of the parameters in terms of powers of the distance along the wall in the flow direction. Hence, the two proposed alternative scaling parameter sets are not necessarily different and distinct. The Falkner-Skan based approach is merely imposing constraints on the functional behavior of the $\delta_s = \mu_1$ and $u_s = u_e$ scaling parameters.

Finally, an important consideration regarding the thickness and shape of the second derivative profile using the moment method is that the observed Gaussian-like behavior is true only for the ZPG turbulent boundary layer. For the APG and FPG cases, it has been observed that the second derivative of the velocity can take on negative values in the neighborhood of the wall.[22] This is, of course, unlike standard probability density function behavior. It is not clear what this means in terms of the behavior for $\mu_1$ and/or $\sigma_v$. There are not many high-quality APG and FPG experimental data sets for which the second derivatives of the velocity can be examined in detail. More research needs to be done to understand the behavior of the second derivative profiles for these cases.

## 5. Conclusions

An alternative to the fixed location sublayer model of the inner region of the turbulent boundary layer is demonstrated. The new model uses readily available experimental data to actually measure the thickness and shape of the inner region of the wall-bounded turbulent boundary layer. Examples are offered to show how the new description method is useful in exploring the actual physics of the boundary layer. Turbulent boundary layer datasets are used to show that this newly defined viscous region includes the Logarithmic Law region. Based on an equal area similarity argument, it is demonstrated that the proposed length scale must be a similarity scaling parameter for the turbulent flow boundary layer.

## Acknowledgment

The author acknowledges the support of the Air Force Research Laboratory and Gernot Pomrenke at AFOSR. In addition, the author thanks the various experimentalists for making their datasets available for inclusion in this manuscript.

## Appendix: Falkner-Skan Properties of the new scaling

In an earlier paper, Weyburne[4] showed that the Prandtl Plus scaling's do NOT behave as similarity scaling parameters for general Falkner-Skan turbulent boundary layers. This brings up the question as to how the new scaling combination $\mu_1$ and $u_e$ fares in comparison. To answer this, we reapply the flow governing equation argument used by Weyburne[4] to this new parameter set. We start off with the Prandtl boundary layer approximation for the *x*-component of the momentum balance given by

$$\overline{u}\frac{\partial \overline{u}}{\partial x} + \overline{v}\frac{\partial \overline{u}}{\partial y} + \frac{\partial}{\partial x}\left\{\overline{\hat{u}^2} - \overline{\hat{v}^2}\right\} + \frac{\partial}{\partial y}\left\{\overline{\hat{u}\hat{v}}\right\} \cong -\frac{1}{\rho}\frac{\partial p}{\partial x} + \nu\frac{\partial^2 \overline{u}}{\partial y^2} \quad , \tag{A.1}$$

where we have used the standard Reynolds decomposition to express the velocities in terms of the average velocities $\overline{u}$ and $\overline{v}$ and the fluctuating components as $\hat{u}$ and $\hat{v}$. We assume that a stream function $\psi(x,y)$ exists such that

$$\psi(x,y) = \delta_s(x) u_s(x) f(\eta) \quad , \tag{A.2}$$

where $f(\eta)$ is a dimensionless function, $\delta_s$ is the length scaling parameter, and $u_s$ is the velocity scaling parameter. For similarity, $f(\eta)$ can only depend on the scaled *y*-position, $\eta$, given by

$$\eta = \frac{y}{\delta_s(x)} \quad . \tag{A.3}$$

The stream function must satisfy the conditions

$$u(x,y) = \frac{\partial \psi(x,y)}{\partial y}, \qquad v(x,y) = -\frac{\partial \psi(x,y)}{\partial x} \quad . \tag{A.4}$$

If we assume the Bernoulli's equation applies to the pressure gradient term, then the stream function version of Eq. A.1 is

$$f''' + \alpha f f'' + \beta\left(1 - f'^2\right) + \left\{\text{reduced Reynolds stress terms}\right\} = 0 \quad , \tag{A.5}$$

where the primes indicate differentiation with respect to $\eta$ and

$$\alpha = \frac{\delta_s^2}{\nu}\frac{du_s}{dx} + \frac{\delta_s u_s}{\nu}\frac{d\delta_s}{dx} \quad \text{and} \quad \beta = \frac{\delta_s^2}{\nu}\frac{du_s}{dx} \quad . \tag{A.6}$$

Eq. A.5 is the Falkner-Skan version of the momentum equation for turbulent flows and the $\alpha$ and $\beta$ terms are the same as the original Falkner-Skan[20] expressions. Similarity of the velocity profile from station to station requires that $\alpha$, $\beta$, and the terms due to the reduced Reynolds stress terms must be constant. The Reynolds Stress terms add additional constraints but these additional constraints do not change the fact that $\alpha$ and $\beta$ must be constant.

We are now in a position to test the similarity scaling set $\mu_1$ and $u_e$. If we take the velocity scaling parameter as

$$u_s(x) = u_e(x) \quad , \tag{A.7}$$

and the length scale is given by

$$\delta_s(x) \; = \; \mu_1(x) \; = \; \frac{\nu u_e(x)}{u_\tau^2(x)} \quad , \tag{A.8}$$

then the parameter $\beta$ is given by

$$\beta \; = \; \frac{\delta_s^2}{\nu}\frac{du_s}{dx} \; = \; \frac{\nu u_e^2}{u_\tau^4}\frac{du_e}{dx} \quad . \tag{A.9}$$

Similarity requires that the parameters $\alpha$ and $\beta$ must be constant from station to station. If we assume analytical functions of the type

$$u_e(x) = \; a_e(x-x_0)^m \quad \text{and} \quad u_\tau(x) = a_\tau(x-x_0)^p \quad , \tag{A.10}$$

where $a_e$, $a_\tau$, $x_0$, *m*, and *p* are constants, then the $\beta$ term becomes

$$\beta \; = \; \frac{\nu u_e^2}{u_\tau^4}\frac{du_e}{dx} \; = \; \frac{\nu m a_e^2}{a_\tau^4}(x-x_0)^{3m-4p-1} \quad . \tag{A.11}$$

Similarity requires $\beta$ is a constant which means that 3*m*-4*p*-1 must be equal to zero (this insures $\alpha$, which in this case is no-zero, is also a constant). Thus, for flows obeying Falkner-Skan power type behavior, the new length and velocity scales are given by

$$u_e(x) = \; a_e(x-x_0)^m \quad , \quad \mu_1(x) = \frac{\nu a_e}{a_\tau^2}(x-x_0)^{(1-m)/2} \quad , \tag{A.12}$$

$$\text{and} \qquad u_\tau(x) = a_\tau(x-x_0)^{\frac{3m-1}{4}} \quad ,$$

which are the same as Falkner-Skan's original expressions.[20] This result indicates that the scaling combination $\mu_1$ and $u_e$ works as similarity scaling parameters for the general Falkner-Skan turbulent boundary layer flow case. In contrast, Weyburne[4] showed that showed this same argument applied to the Prandtl Plus scaling's results in $\alpha = 0$ and $\beta \propto 1/x$ which are defining characteristics of sink flows.

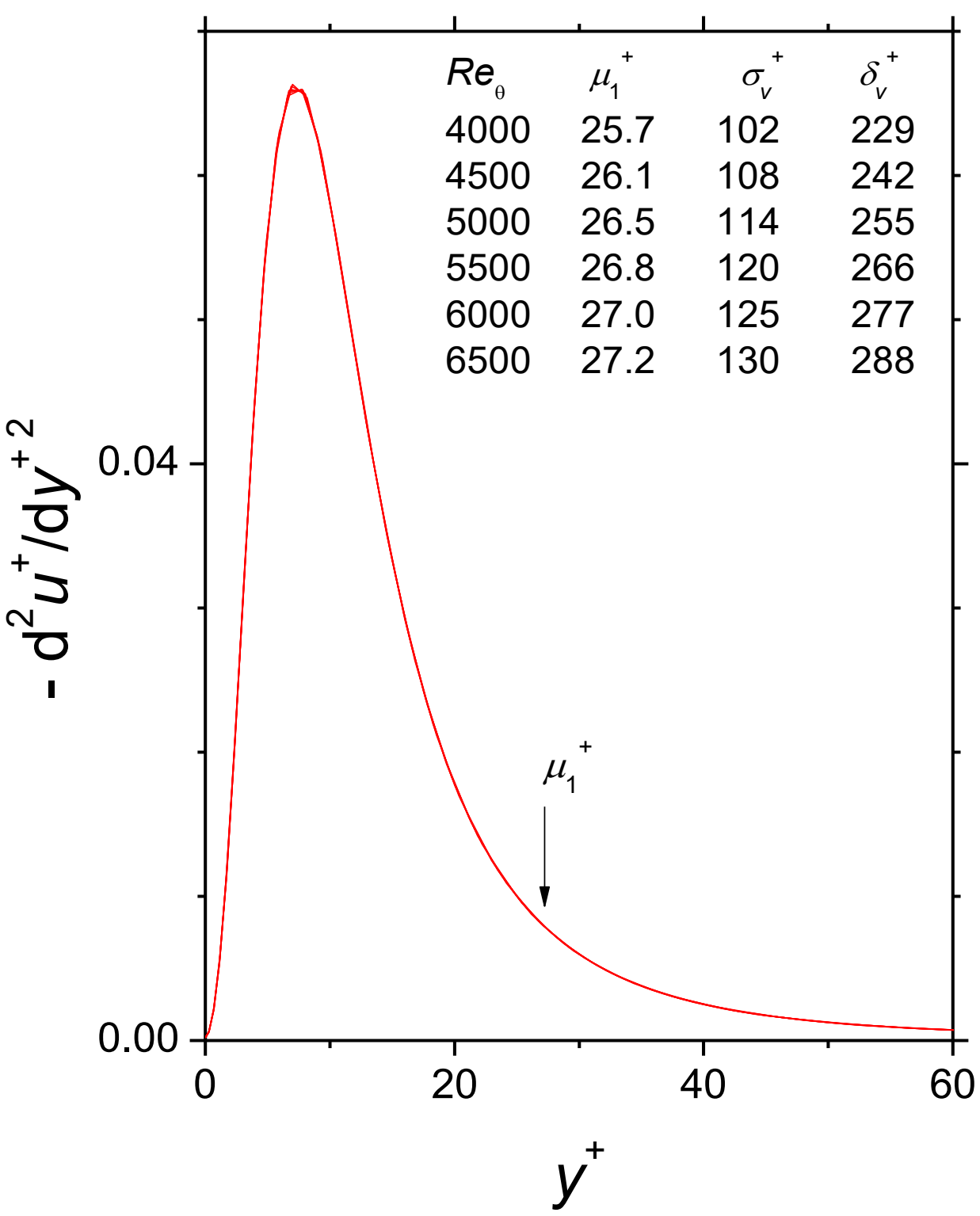

FIG. 1. Six second derivative velocity profile curves from Sillero, Jimenez, and Moser[10] plotted in Plus units.

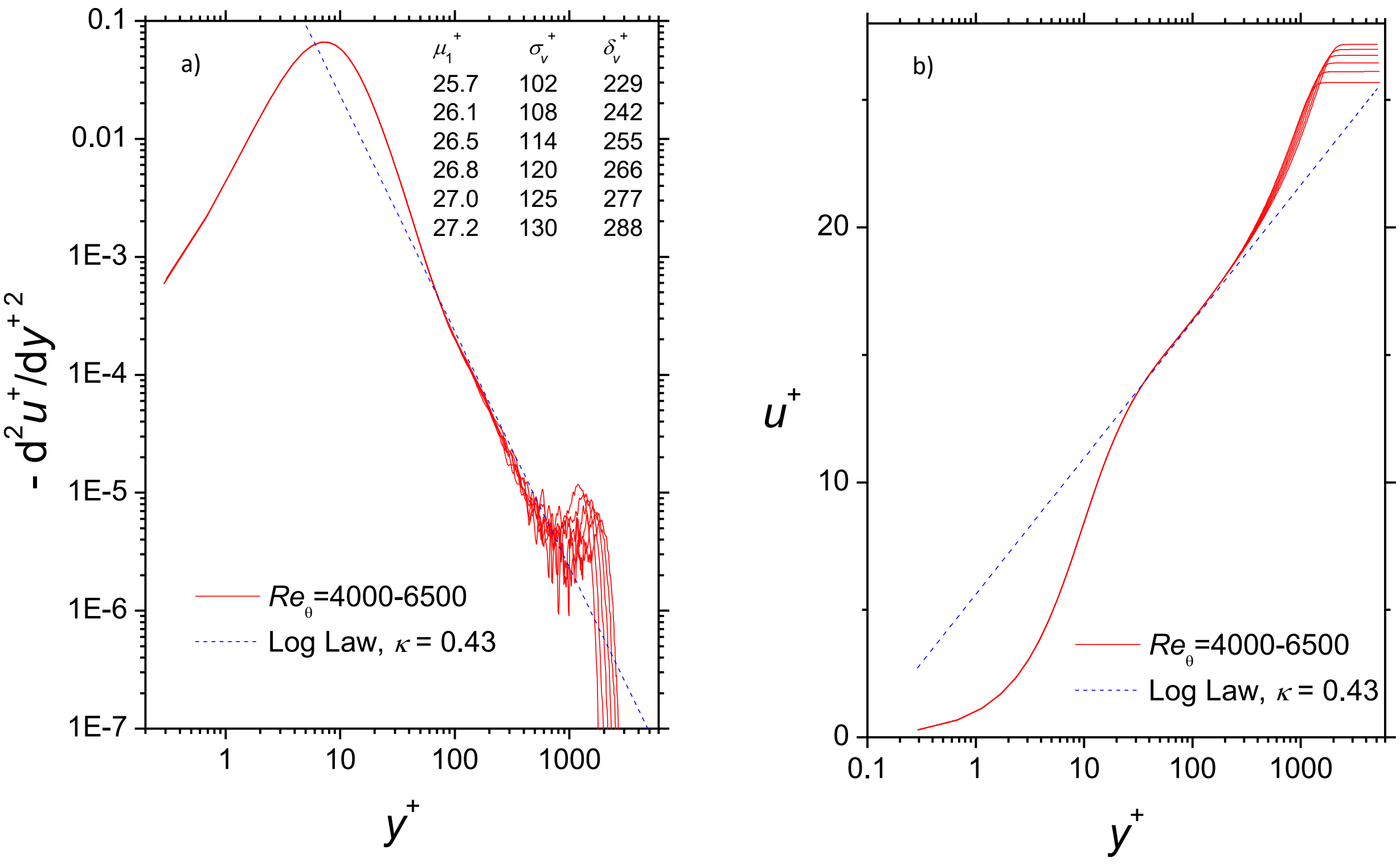

Figure 2: The solid lines are the six Sillero, Jimenez, and Moser[10] a) second derivative profiles and b) velocity profiles plotted in plus units.

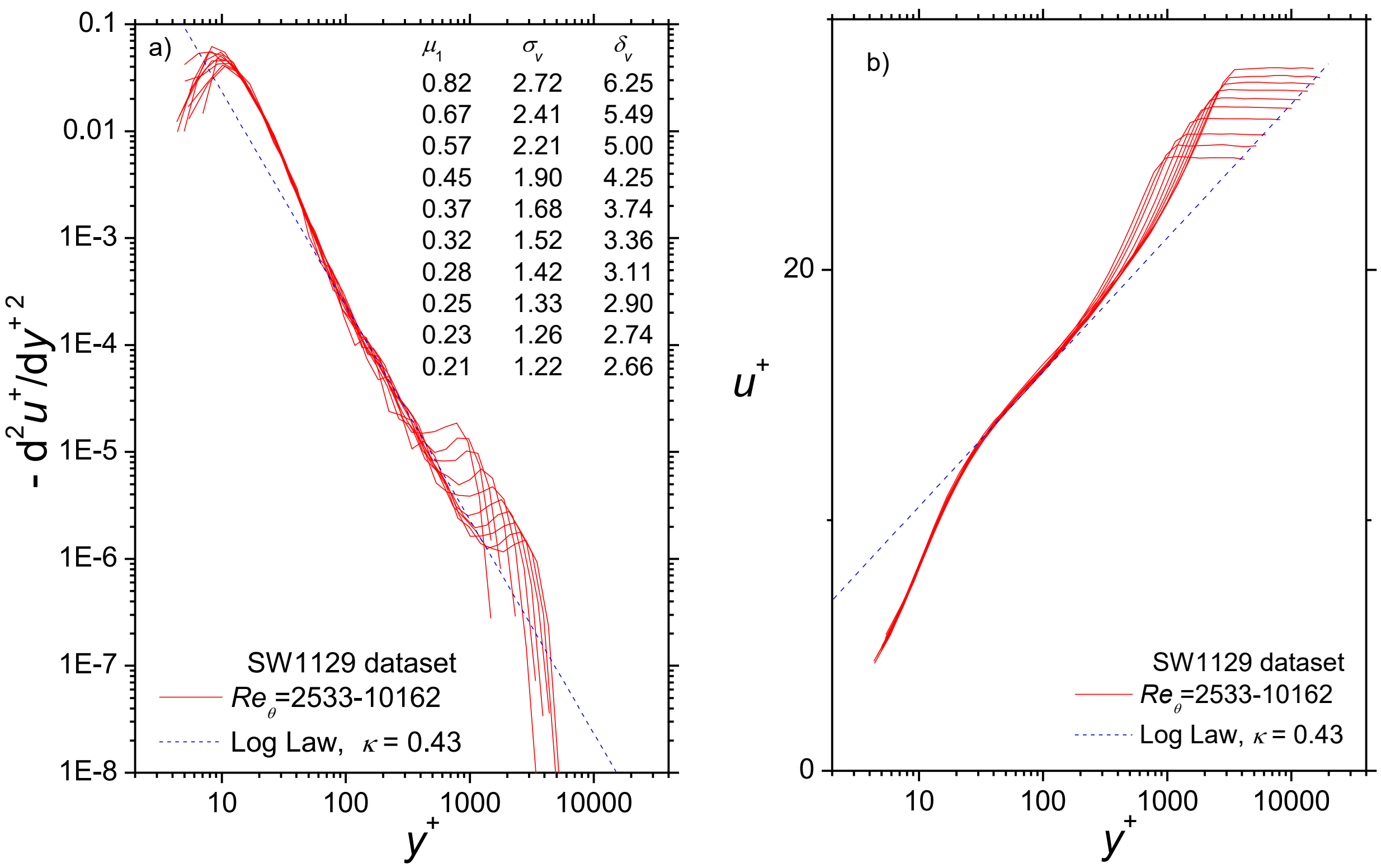

Figure 3: The solid lines are the Österlund[11] a) second derivative profiles and b) velocity profiles plotted in plus units. The inset units are all in millimeters.

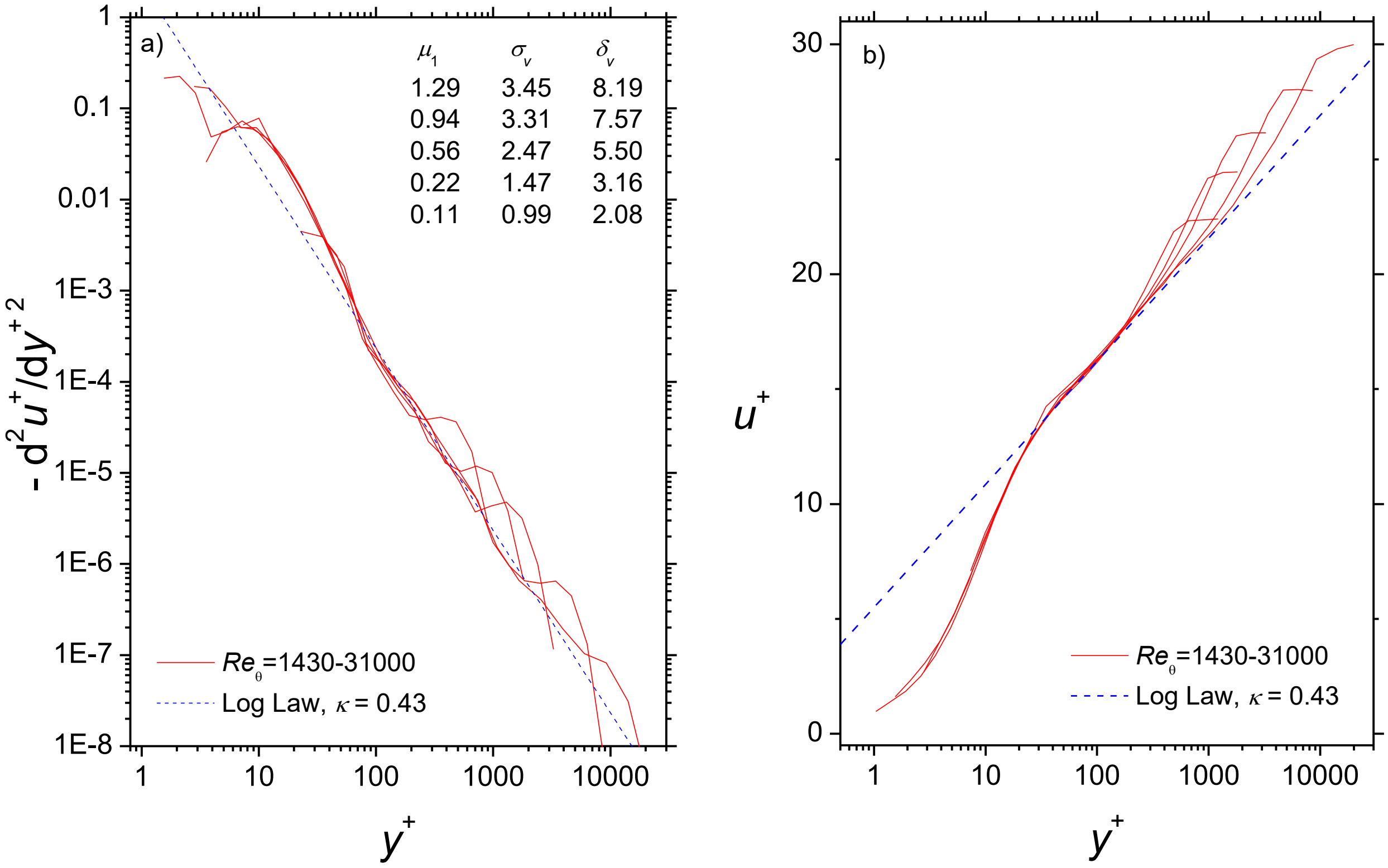

Figure 4. The solid lines are the DeGraaff and Eaton[12] a) second derivative profiles and b) velocity profiles plotted in plus units. The inset units are all in millimeters.

## Tables

Table 1: Data from Wieghardt and Tillmann[17] in millimeters.

| $Re_\theta$ | $\mu_1$ | $\sigma_v$ | $\delta_v$ |
|---|---|---|---|
| 457.5 | 0.171 | 0.35 | 0.87 |
| 1067.8 | 0.217 | 0.55 | 1.32 |
| 1558.7 | 0.237 | 0.68 | 1.60 |
| 2007.6 | 0.253 | 0.80 | 1.85 |
| 2397.8 | 0.265 | 0.89 | 2.04 |
| 2864.3 | 0.277 | 0.98 | 2.24 |
| 3436.6 | 0.294 | 1.11 | 2.52 |
| 3856.3 | 0.295 | 1.17 | 2.64 |
| 4387.2 | 0.297 | 1.26 | 2.81 |
| 4858.4 | 0.304 | 1.34 | 2.98 |
| 5473.1 | 0.313 | 1.44 | 3.20 |
| 6228.9 | 0.323 | 1.56 | 3.45 |
| 7170.3 | 0.420 | 1.89 | 4.21 |
| 8172.4 | 0.340 | 1.83 | 3.99 |
| 8896.5 | 0.345 | 1.92 | 4.18 |
| 9814.5 | 0.352 | 2.03 | 4.41 |
| 10611.2 | 0.352 | 2.11 | 4.56 |
| 11471.6 | 0.357 | 2.21 | 4.77 |
| 12222.6 | 0.362 | 2.29 | 4.94 |
| 13042.8 | 0.380 | 2.43 | 5.24 |
| 14024.4 | 0.380 | 2.51 | 5.40 |
| 14703.2 | 0.371 | 2.53 | 5.43 |
| 15517.7 | 0.377 | 2.62 | 5.62 |

Table 2: Data from Smith[18] in millimeters.

| $Re_\theta$ | $\mu_1$ | $\sigma_v$ | $\delta_v$ |
|---|---|---|---|
| 4601 | 0.316 | 1.33 | 2.97 |
| 4980 | 0.319 | 1.39 | 3.10 |
| 5388 | 0.328 | 1.47 | 3.27 |
| 5888 | 0.330 | 1.53 | 3.39 |
| 6866 | 0.354 | 1.73 | 3.82 |
| 7696 | 0.358 | 1.84 | 4.04 |
| 9148 | 0.368 | 2.03 | 4.44 |
| 10347 | 0.378 | 2.19 | 4.77 |
| 11608 | 0.382 | 2.30 | 4.99 |
| 13189 | 0.389 | 2.50 | 5.39 |